# Graphene-Based Couplers: A Brief Review

**Mohammad Bagher Heydari [1,*], Mohammad Hashem Vadjed Samiei [1]**

[1] School of Electrical Engineering, Iran University of Science and Technology (IUST), Tehran, Iran

*Corresponding author: mo_heydari@elec.iust.ac.ir

**Abstract:** Graphene is an interesting debated topic between scientists because of its unique properties such as tunable conductivity. Graphene conductivity can be varied by either electrostatic or magnetostatic gating or via chemical doping, which leads to the design of various photonic and electronic devices. Among various graphene-based structures, plasmonic graphene couplers have attracted the attention of many researchers because of their fascinating applications in the THz frequencies. There are four main types of graphene couplers proposed in the literature, which are: 1- directional, 2- non-reciprocal, 3- dielectric, and 4- nano-ribbon couplers. This paper aims to study and investigate the various types of graphene-based couplers published in the literature.

## 1. Introduction

Plasmonics is a new emerging science in recent years, which has many fascinating properties in photonics and electronics [1, 2]. Both metals and two-dimensional (2D) materials can support surface plasmon polaritons (SPPs). In metals, SPPs often are excited in the visible and near-infrared regions [3, 4]. The emerging 2D materials like molybdenum disulfide ($MoS_2$) and tungsten diselenide ($WSe_2$) have generated immense interest for semiconductor and nanotechnology due to their fascinating properties [5]. Among these 2D materials, carbon nanotubes exhibit attractive features that are widely used in chemistry and physics [6, 7]. Graphene is a 2D sheet that offers several fundamentally superior properties that has the potential to enable a new generation of technologies [8-10].

The graphene research has grown slowly in the late 20th century but AB initio calculations illustrated that a single graphene layer is unstable [11]. In 2007, Andre Geim and Konstantin Novoselov first isolated single layer samples from graphite, and the Nobel prize in physics 2010 was awarded to them [12, 13].

Graphene is one of the allotropes of carbon, which is a planar monolayer of carbon atoms that form a honeycomb lattice with a carbon-carbon bond length of 0.142 nm [14]. It exhibits a large wide of interesting properties [15-20]. One of them is its highly unusual nature of charge carriers which behave as Dirac fermions [21]. This feature has a great effect on the energy spectrum of Landau levels produced in presence of a magnetic field [22, 23]. The Hall conductivity is observed at zero energy level. Also, the Hall Effect is distinctly different than the conventional Hall Effect which is quantized by a half-integer [24, 25]. Graphene absorbs 2.3% of incident light and the absorption can be tuned by varying the Fermi level through the electrical gating [26, 27]. This property is widely used in designing transparent electrodes in solar cells [28-30]. Graphene has an exceptional thermal conductivity (5000 w/m$^{-1}$ K$^{-1}$) that is utilized in fabricating and designing the electronic components [31].

Graphene transport characteristics and electrical conductivity can be tuned by either electrostatic or magnetostatic gating or via chemical doping, which this conductivity leads to the design of various photonic and electronic devices [2, 32-38]. All of the mentioned fascinating features makes the graphene as a good candidate for designing novel devices in the THz frequencies such as waveguides [39-45], filters [46], Radar Cross-Section (RCS) reduction-based devices [47-49], and graphene-based medical devices [50-56]. Furthermore, integration graphene with anisotropic materials such as ferrites, which have many interesting properties in the microwave regime [57, 58], can enhance the performance of the graphene-based devices [40]. One of the interesting graphene-based devices designed by graphene conductivity is the plasmonic coupler which has various applications in communications. This paper presents a short review of plasmonic graphene couplers.

The paper is organized as follows. Due to the importance of graphene conductivity in designing plasmonic couplers, the conductivity of graphene in presence of electrostatic or magnetostatic bias has been considered in section 2. Then, various types of graphene couplers are introduced and their performance principles are discussed in section

3. These couplers can be categorized into four groups: 1-directional couplers, 2- non-reciprocal couplers, 3- dielectric couplers, and 4- nano-ribbon couplers. In all of these couplers except non-reciprocal couplers, the graphene has been biased electrically while the anisotropic graphene has been used in designing the non-reciprocal couplers. Section 4 briefly considers the perspective and challenges. Finally, section 5 concludes the paper.

## 2. The Conductivity of Graphene

In graphene Plasmonics, the conductivity of graphene plays the main role in designing photonic devices, because it describes the electromagnetic interactions between the graphene sheet and the external field. The popular relation used for the conductivity has been computed via Kubo's formula [59]:

$$\sigma(\omega,\mu_c,\Gamma,T) = \frac{je^2(\omega-j2\Gamma)}{\pi\hbar^2}\left[\frac{1}{(\omega-j2\Gamma)^2}\int_0^\infty \varepsilon\left(\frac{\partial f_d(\varepsilon)}{\partial \varepsilon}-\frac{\partial f_d(-\varepsilon)}{\partial \varepsilon}\right)d\varepsilon - \int_0^\infty \frac{f_d(\varepsilon)-f_d(-\varepsilon)}{(\omega-j2\Gamma)^2-4(\varepsilon/\hbar)^2}\left(\frac{\partial f_d(\varepsilon)}{\partial \varepsilon}-\frac{\partial f_d(-\varepsilon)}{\partial \varepsilon}\right)d\varepsilon\right] \quad (1)$$

Where

$$f_d(\varepsilon) = \frac{1}{1+\exp\left(\frac{\varepsilon-\mu_c}{K_BT}\right)} \quad (2)$$

In the above equations, $e$ is the charge of an electron, $\hbar$ is the reduced Planck's constant, $K_B$ is Boltzmann's constant, $T$ is temperature, $\omega$ is radian frequency, and $\mu_c$ is chemical potential. By substituting (2) into (1) and integrating, one can obtain [59]:

$$\sigma(\omega,\mu_c,\Gamma,T) = \sigma_{\text{inter}}(\omega) + \sigma_{\text{intra}}(\omega) = \frac{-je^2}{4\pi\hbar}Ln\left[\frac{2|\mu_c|-(\omega-j2\Gamma)\hbar}{2|\mu_c|+(\omega-j2\Gamma)\hbar}\right] + \frac{-je^2 K_B T}{\pi\hbar^2(\omega-j2\Gamma)}\left[\frac{\mu_c}{K_BT}+2Ln\left(1+e^{-\mu_c/K_BT}\right)\right] \quad (3)$$

In (3), the conductivity has been split into two terms: intra-band and inter-band transitions. It should be noted that the intra-band electronic process becomes dominant in the mid-infrared and THz frequencies at room temperature. All of the plasmonic graphene couplers utilize the relation (3) for their design except non-reciprocal couplers. In the existence of the external magnetic field, the conductivity of graphene converts to tensor [60]:

$$\bar{\bar{\sigma}}(\omega,\mu_c,\Gamma,T,\vec{B}_0) = \begin{pmatrix} \sigma_d & -\sigma_o \\ \sigma_o & \sigma_d \end{pmatrix} \quad (4)$$

Where $\sigma_d, \sigma_o$ are [60]:

$$\sigma_d(\mu_c,B_0) = \frac{e^2 v_f^2 |eB_0|(\omega-j2\Gamma)\hbar}{-j\pi} \times$$

$$\sum_{n=0}^{\infty}\left[\frac{f_d(M_n)-f_d(M_{n+1})+f_d(-M_{n+1})-f_d(-M_n)}{(M_{n+1}-M_n)^2-(\omega-j2\Gamma)^2\hbar^2}\times\left(1-\frac{\Delta^2}{M_n M_{n+1}}\right)\times\frac{1}{M_{n+1}-M_n}\right. \quad (5)$$

$$\left.+\frac{f_d(-M_n)-f_d(M_{n+1})+f_d(-M_{n+1})-f_d(M_n)}{(M_{n+1}+M_n)^2-(\omega-j2\Gamma)^2\hbar^2}\times\left(1+\frac{\Delta^2}{M_n M_{n+1}}\right)\times\frac{1}{M_{n+1}+M_n}\right]$$

$$\sigma_o(\mu_c, B_0) = \frac{e^2 v_f^2 |eB_0|}{\pi} \times$$

$$\sum_{n=0}^{\infty} \left[ \frac{f_d(M_n) - f_d(M_{n+1}) - f_d(-M_{n+1}) + f_d(-M_n)}{(M_{n+1} - M_n)^2 - (\omega - j2\Gamma)^2 \hbar^2} \times \left(1 - \frac{\Delta^2}{M_n M_{n+1}}\right) \right]$$
$$+ \left[ \frac{f_d(M_n) - f_d(M_{n+1}) - f_d(-M_{n+1}) + f_d(-M_n)}{(M_{n+1} + M_n)^2 - (\omega - j2\Gamma)^2 \hbar^2} \times \left(1 + \frac{\Delta^2}{M_n M_{n+1}}\right) \right]$$
(6)

In the above relations, $f_d$ has been defined in (2) and,

$$M_n = \sqrt{\Delta^2 + 2n v_f^2 |eB_0| \hbar} \tag{7}$$

In (7), $v_f \approx 10^6 \, m/s$ is the electron velocity in graphene, $B_0$ is applied magnetic field, and $\Delta$ is an excitonic energy gap. Relations (5) and (6) become simple for $\mu_c \gg \hbar \omega_c$ and $\mu_c \gg K_B T$ [61]:

$$\sigma_d(\omega, \mu_c, \tau, T, B_0) = \sigma_0 \frac{1 + j\omega\tau}{(\omega_c \tau)^2 + (1 + j\omega\tau)^2} \tag{8}$$

$$\sigma_o(\omega, \mu_c, \tau, T, B_0) = \sigma_0 \frac{\omega_c \tau}{(\omega_c \tau)^2 + (1 + j\omega\tau)^2} \tag{9}$$

Where

$$\omega_c = \frac{eB_0 v_f^2}{|\mu_c|} \tag{10}$$

is cyclotron frequency and the static conductivity for $B_0 = 0$ is expressed as [61]:

$$\sigma_0 = \frac{e^2 \mu_c \tau}{\pi \hbar^2} \tag{11}$$

In (11), $\tau$ is scattering time which is defined as [61]:

$$\tau = \frac{\pi \hbar^2 n_s \mu}{e \mu_c} \tag{12}$$

Where $\mu$ is the DC mobility of graphene and $n_s$ is the surface carrier density.

## 3. Various Types of Plasmonic Graphene Couplers

In the literature, various types of graphene couplers have been proposed. Here, we categorize them into four main groups, as illustrated in Fig. 1. This section tends to study each kind of graphene coupler in a separate sub-section.

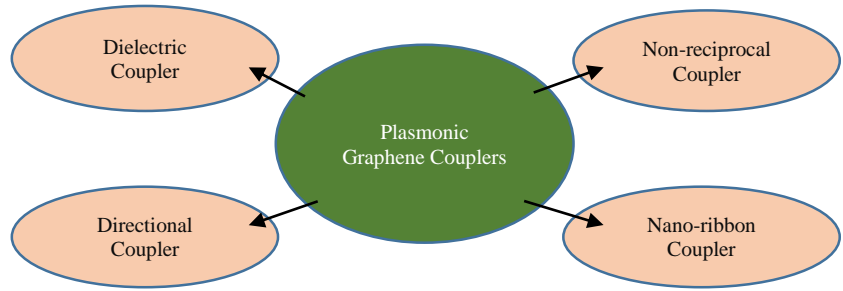

Fig 1. Various types of plasmonic graphene couplers

*3.1 Directional Couplers*

Directional couplers are passive devices that are utilized in the microwave and THz Technology. They usually are composed of two coupled transmission lines where are close enough together. The electromagnetic power in a transmission line is coupled to another line or port. They are called "directional" because they can couple the electromagnetic power only in one direction. They have many applications such as feedback, antenna beam-forming, and separating various signals in telephone lines. In the THz regime, graphene-based directional couplers with good performance characteristics have been introduced and investigated.

Hong Ju Li et al. have proposed directional couplers based on graphene in 2013 [62]. They used the FDTD method to simulate the proposed structure. The coupler consists of double-layer graphene where are located at a distance of $d$ from each other and have coupling length $L$, as seen in fig 2. The magnetic field distributions $H_z$ for $L=220, 440$ nm and $d=50$ nm have been illustrated in fig. 2. It is clear that by increasing L, low transmission at the output occurs. Then, the authors focus on the switch effect which can be used in nano-integrated circuits [62]. In [62], three-layer graphene couplers have been simulated and fabricated, which acts as a directional coupler.

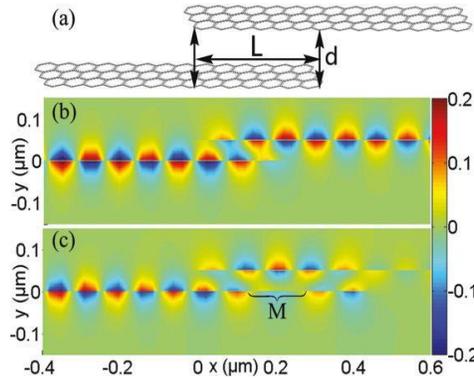

Fig. 2. (a) The double-layer graphene-based coupler; Magnetic field distribution $H_z$ for (b) *L=220 nm*, (c) *L=420 nm* [62]

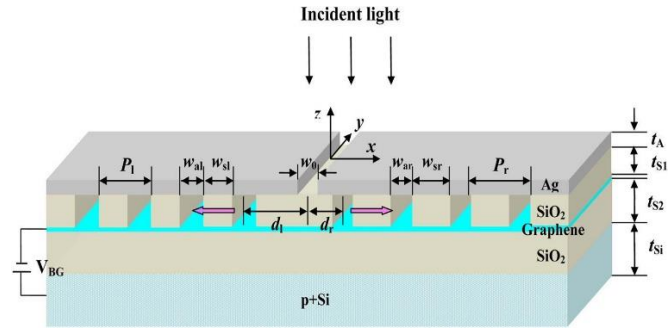

Fig. 3. The schematic of graphene directional coupler designed in [63]

In [63], the graphene-based tunable plasmonic directional coupler is reported. This coupler consists of the dielectric grating, a graphene sheet, a thin metal film, and a dielectric film, as illustrated in fig 3 [63]. Indeed, this directional coupler is a hybrid coupler based on metal-graphene. The $SiO_2$-Graphene-$SiO_2$ is a Bragg reflector which its band-gap can be tuned by graphene conductivity[63]. A plane wave with TM polarization illuminates the whole structure. The authors showed that this structure can be used as the unidirectional coupler and switching of SPPS in Bragg Reflector can be occurred due to varying chemical potential [63]. An electrically controllable directional coupler is studied in [64]. The structure is composed of two parallel identical straight dielectric-loaded graphene plasmonic waveguide where are separated by d and are connected to two S-shape bends, as exhibited in fig. 4(a). It should be noted that maximal power coupled from input to cross waveguide can be controlled by the chemical potential of graphene [64]. So, this device can be used as a switch in which authors have reported an extinction ratio larger than 16 dB for it [64].

One of the applications of the directional coupler is using them to design novel devices such as switch and the beam splitter. For instance, the authors in [65] have investigated the compact polarization beam splitter based on graphene asymmetrical directional coupler. The splitter has been constructed by silicon waveguide and graphene multilayer embedded silicon waveguide [65]. The authors have achieved an extinction ratio of 21.2 dB and low insertion loss of 0.36 dB for 8.3 $\mu m$-long coupler and 200 nm wide gap separation [65].

Thang Q.Tran and Sangin Kim introduced a controllable vertical directional coupler which is based on Graphene-assisted frustrated total internal reflection (GA-FTIR) [66]. Fig. 4(b) displays the vertical directional coupler which consists of two curved Si waveguides separated by graphene-SiO$_2$-graphene layers [66]. The coupler works with changing the chemical potential of graphene, where controls the coupling length of the directional coupler [66]. The authors concluded that this is a compact device with very low losses ($\approx 0.11\,dB$) for both blocking and coupling states and a strong extinction ratio of 24 dB [66]. In [67], the graphene directional coupler is presented based on two dielectric ridges. In this coupler, guided TM polarized mode is propagated in each ridge and by changing the chemical potential of graphene, the coupler can vary its state from bar state to cross state [67]. It should be noted that this coupler support even and odd supermodes due to hybridization of two modes [67]. In the rest of the paper, the authors discuss the beat length of the coupler and display the field distribution of the proposed directional coupler in different chemical potentials [67].

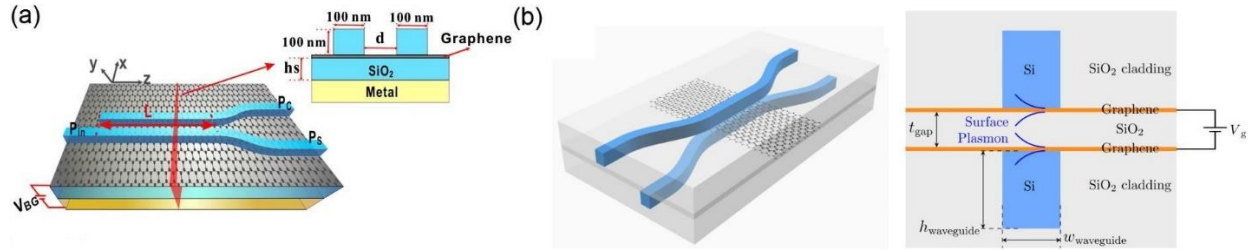

Fig. 4. (a) The controllable directional coupler based on graphene [64],
(b) The vertically directional coupler where two Si layer are separated by graphene-SiO$_2$-graphene layers [66]

*3.2 Non-reciprocal couplers*

As mentioned before, when graphene is magnetically biased, the graphene conductivity becomes tensor (see relations (8) and (9)) and two edge modes can be propagated at edges with different dispersions. A non-reciprocal coupler by using an anisotropic graphene sheet has been introduced by Nima Chamanara et al. in 2013 [68] and developed work for the application of this switchable graphene coupler as a magnetic probe is reported in [69]. The authors are applied edge modes to design a non-reciprocal coupler which is composed of two parallel plasmonic waveguides with anisotropic graphene, as shown in fig. 5 [68]. Dispersion curves for bulk and edge modes for two separate graphene strips with different carrier densities are represented in fig. 6 [68].

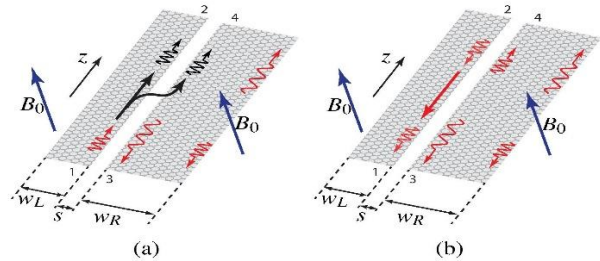

Fig. 5. (a) Non-reciprocal graphene coupler, (a) Feeding through port 1, (b) Feeding through port 2[68]

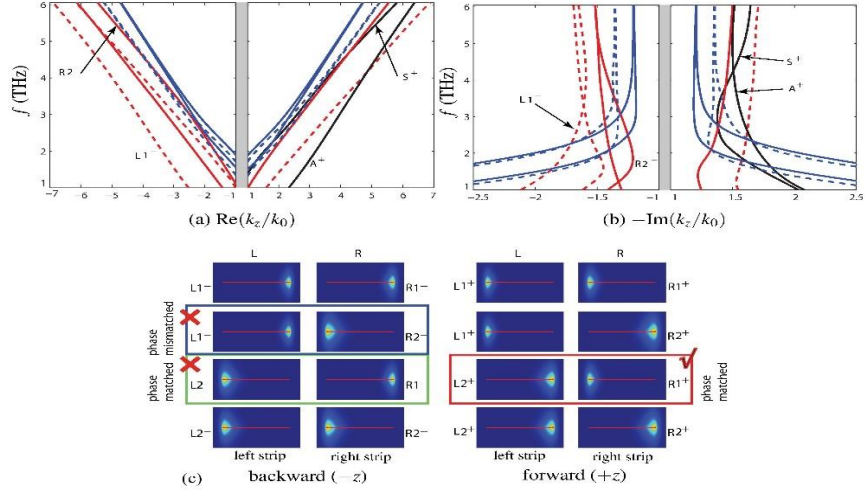

Fig. 6. (a), (b) Dispersion curves for two isolated graphene strip for $w = 100\,\mu m$, $B_0 = 1T$ and $\tau = 0.1\,ps$. The solid (right strip) and dashed (left strip) curves are graphene strips with $n_s = 10^{13}\,cm^{-2}, 8\times 10^{12}\,cm^{-2}$, respectively. The red and blue lines represent edge and bulk modes and phase-matched regions are displayed by ellipses, (c) electric field distribution of edge modes for strip represented in fig. 5 (a),(b) [68]

It is obvious from fig. 6 that mode R1+ propagates on the left edge of the right strip and mode L2+ propagates on the right edge of the left strip and these modes can be coupled due to phase-matched condition [68]. But for backward direction (-z), the matching wave number R1- and L2- are on the opposite edges of two strips, and also R2- and L1- cannot be coupled due to different wave number [68]. Therefore, as feed port 2 selected for backward direction, no coupling occurs where is exhibited in fig. 6(c) [68]. Paper [69] has a similar context as [68] but it designs the coupler as a magnetic probe. This magnetic sensor is constructed by two broadside coupler graphene strips connected by a DC source [69]. The details are explained in this paper and the reader can refer to it [69]. In [70], unidirectional surface plasmons in non-reciprocal graphene-based gyrotropic interfaces are considered in detail. Authors of this paper have investigated a four-port graphene-based SPP coupler where $H_z$ field distributions of this structure for different magnetic bias and chemical potentials have been demonstrated in the following figure [70]. By properly selecting $B_0$ and $\mu_c$, SPP waves from port 1 can be coupled to the other three ports. For instance, for case (d), with $B_0=0$ and $\mu_c = 0.096\,eV$, the device operates as a backward coupler [70].

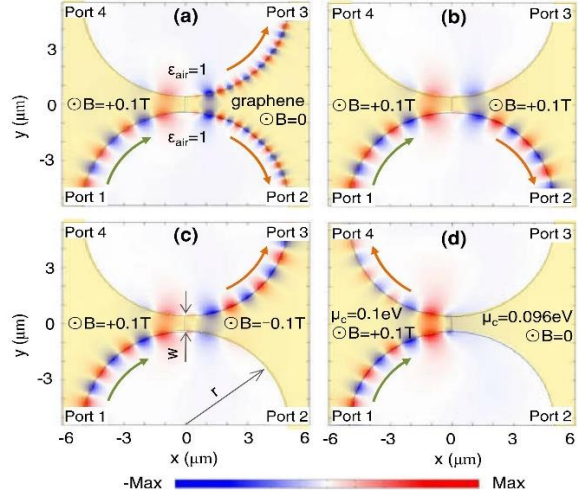

Fig. 7. Magnetic field distributions ($H_z$) of four-port non-reciprocal graphene coupler at $f=42.7\,THz$ for geometrical parameters of $r = 5\,\mu m$, $w = 0.8\,\mu m$ and $\mu_c = 0.1\,eV$ when the magnetic field at the right side is: (a) $B_0 = 0$, (b) $B_0 = +0.1T$, (c) $B_0 = -0.1T$ and (d) $B_0 = 0$ with different chemical potential of $\mu_c = 0.096\,eV$ [70]

*3.3 Dielectric couplers*

In this section, we review graphene-based planar couplers with dielectric layers. Bing Wang et al. have considered the coupling of TM SPPs between two graphene sheets separated by a dielectric with a thickness of $d=2a$ and permittivity of $\varepsilon_h$ [71]. The authors derived an approximate relation for propagation constant and have simulated the structure by FDTD where the electric field distributions of symmetric and anti-symmetric modes for $d=50\ nm$ are represented in fig. 8 [71]. In this work, the proposed coupler has been utilized to design various devices such as optical splitter and Mach-Zehnder interferometer [71]. In another research article reported by Bing Wang and his co-workers, the coupling of SPPs in monolayer graphene sheet arrays has been focused and the behavior of SPPs for out of the plane and in-plane illuminations are discussed in detail [72].

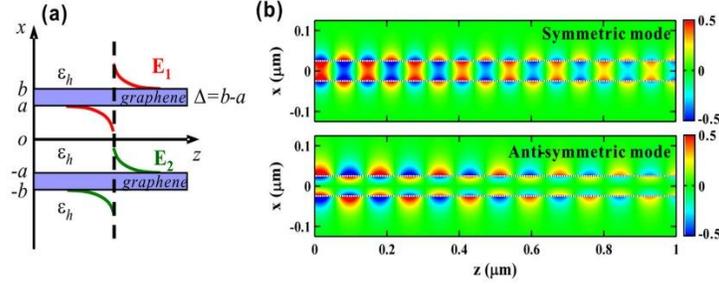

Fig. 8. (a) The graphene coupler with two graphene sheets separated by a dielectric,
(b) Electric field ($E_x$) distributions of symmetric and anti-symmetric modes for $d=50\ nm$ [71]

In [73], a tunable dielectric coupler based on graphene has been designed whose the beat length of the coupler can be controlled by the bias voltage of graphene, as exhibited in fig. 9. This waveguide has $w=4\ \mu m$ and the core and cladding have a relative dielectric constant of 3 and 2.25, respectively. In this work, the variations of the beat length are modeled by an analytical relation and the beat length as a function of chemical potential for different $D$ has been plotted in fig. 9 (b) [73]. It is obvious that the beat length depends on the distance of waveguides [73].

A planar patch plasmonic graphene coupler has been investigated in [74]. This coupler had a different structure from previous dielectric couplers. A 4-port graphene patch has been located on the dielectric layer, as shown in fig. 10. The authors have achieved a semi-analytical relation by starting the HIE-FDTD method and then they have simulated the following patch coupler [74]. The normalized $H_z$ distributions on the graphene coupler are represented where the frequency of incident wave at port 1 are assumed *1.5, 1 THz* [74]. The incident power can be coupled to port 2,3 and 4 when the frequency of the incident wave at port 1 is *1.5 THz* while the incident power is only coupled to port 2 for the frequency is *1 THz* [74]. A parametric study for considering the influence of various parameters such as substrate permittivity on s-parameters of patch graphene has been reported in [74]. The authors have shown that the central frequency bandwidth of $f_0$ can be varied by changing the chemical potential of graphene [74]. This coupler can be utilized as a four-port graphene circulator in THz frequencies [74].

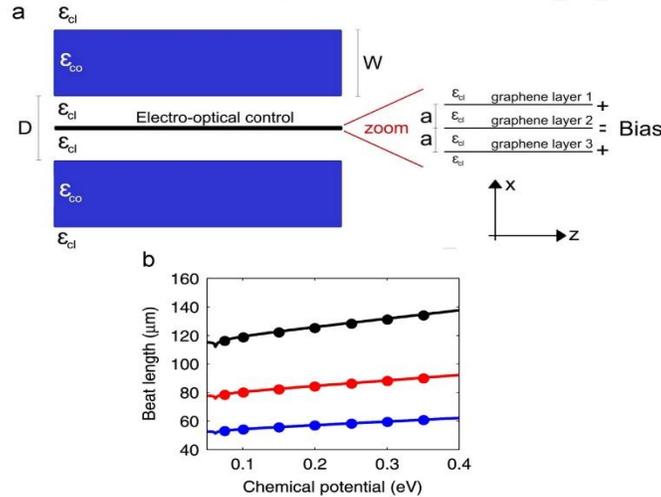

Fig. 9. (a) The schematic of tunable graphene coupler, (b) Beat length as a function of chemical potential for various $D$ ($D=1\ \mu m$(blue), $2\ \mu m$(red), $3\ \mu m$(black)) for $\varepsilon_{co}=3$, $\varepsilon_{ci}=2.25$ and $w=4\ \mu m$ [73]

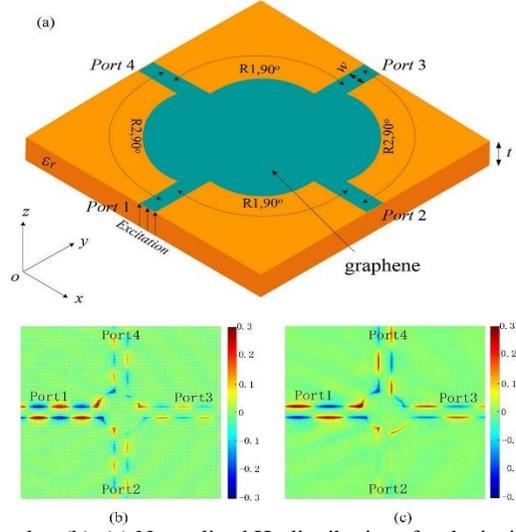

Fig. 10. (a) The patch graphene coupler, (b), (c) Normalized $H_z$ distributions for the incident wave with the frequency of 1.5,1 THz, respectively. The geometrical parameters are supposed $R_1 = 20\,\mu m$, $R_2 = 25\,\mu m$, $w = 12\,\mu m$ and the dielectric constant of the substrate is 2.33 [74]

*3.4 Nano-ribbon couplers*

Graphene Nano-Ribbon (GNR) is a graphene sheet with a finite width in the nano dimension. In this section, graphene nano-ribbon (GNR) couplers have been introduced and studied more precisely. It should be noted that the structure of nano-ribbon couplers is similar to dielectric couplers. Bin Sun et al. investigated TM SPP waves transporting from a graphene infinite layer to a graphene strip with width *w*, as demonstrated in fig. 11 [75]. In this research, the authors firstly investigated the coupling coefficients and coupling length as a function of wavelength for symmetric and anti-symmetric modes [75]. By simulating the structure for various GNR width (*w*), the influence of chemical potential and distance between the strip and infinite graphene layer (*d*) on power transmitted at the output graphene is depicted in fig. 12 [75]. Fig. 12 (a) indicates that the transmission spectrum oscillates with distance (*d*) and also the transmission has a wide dip for *w=130 nm* while it has two dips for *w=200,260 nm* [75]. It is obvious from fig. 12 (b) that there is a blocked area where the input power cannot couple to graphene strip for various chemical potentials [75]. The authors showed that GNRs can block plasmonic waves to transport on the graphene, which occurs due to the forming of standing waves and Fabry-Perot resonance [75].

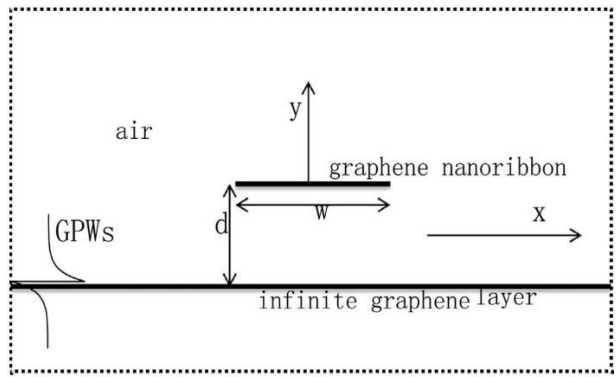

Fig. 11. GNR coupler composed of an infinite layer and a graphene strip [75]

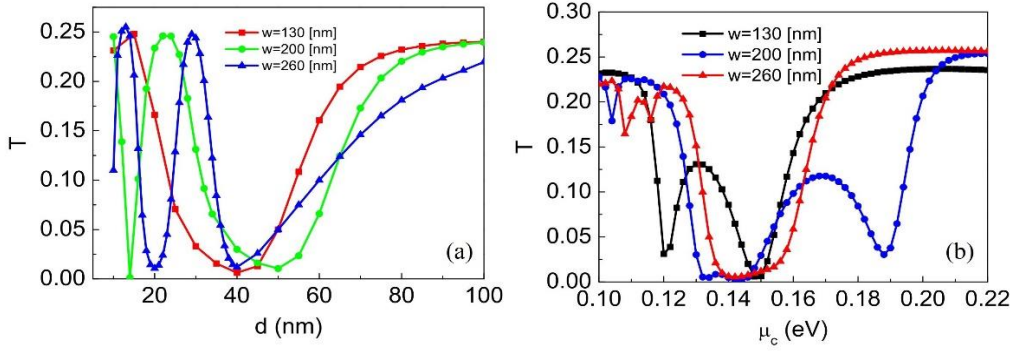

Fig. 12. Transmission spectrum as function of (a) *d*, (b) chemical potential, for various strip widths (*w*) [75]

In [76], two structures for nano couplers based on GNRs have been proposed. The first proposed coupler is illustrated in fig. 13 (a) where a semi-infinite graphene sheet is located at a distance of *d* from the nanoribbon with length *L* [76]. The properties of this coupler are described by applying coupled-mode theory (CMT). The normalized $|H_z|$ field at the wavelength of $\lambda = 7.95\,\mu m$ is exhibited in fig. 13(b) [76]. It indicates that the field is concentrated in the nanoribbon region. In the rest of this paper, a directional plasmonic coupler similar to the structure in fig. 11 has been considered and studied in detail [76].

A coupler designed by two parallel graphene nanostrips arrays is presented in [77]. In the unit cell of the coupler, two coplanar graphene nanostrips are deposited on the dielectric substrate and the whole structure is illuminated by an incident wave, as illustrated in fig. 14 (a) [77]. This coupler also can be categorized in the dielectric graphene coupler group. The periodic boundary conditions are assumed to be a perfectly matched layer (PML) in *x,y* directions and the FDTD method is applied to numerically study the structure [77]. The transmission spectrum versus frequency indicates that the low-energy peak is broadened as parameter *s* increases, while this happens inversely for high-energy one [77]. This structure exhibits Tunable plasmon-induced transparency, which can be utilized in slow light and optical switching [77].

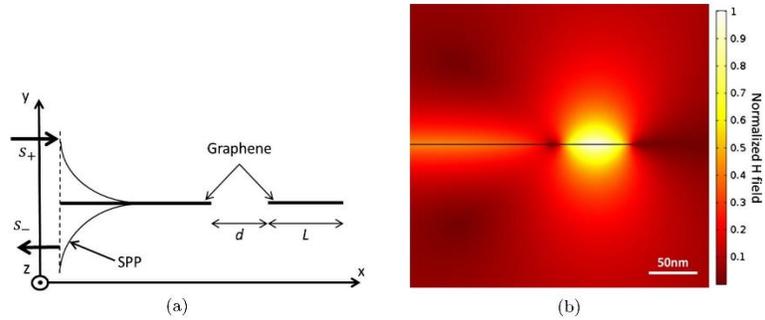

Fig. 13. (a) The schematic of plasmonic coupler based on GNR,
(b) the normalized $|H_z|$ field at $\lambda = 7.95\,\mu m$ for geometrical parameters of *L=75 nm* and *d= 10 nm* [76]

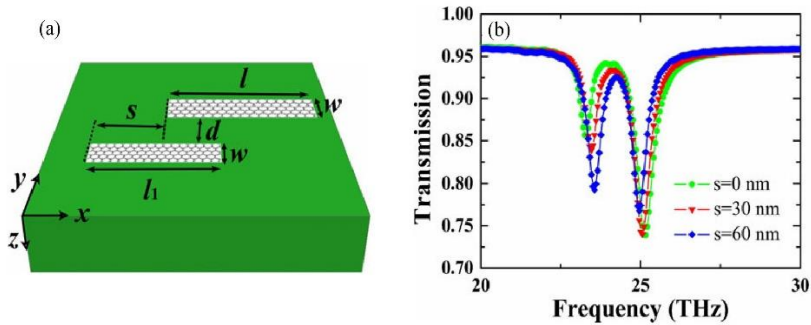

Fig. 14. (a) The unit cell of periodic plasmonic coupler based on two GNRs, (b) Transmission spectrum as a function of frequency for various parameter *s* for geometrical parameters of *w = 30 nm, d = 60 nm, l₁ = 120 nm* and *l= 130 nm* and dielectric constant of 1.5; The periodicity is *450 nm* in both x and y directions [77]

## 4. Perspectives and Challenges

Since graphene couplers are designed based on the optical conductivity of graphene, they provide some useful benefits, which are important in mid-infrared frequencies. The most popular advantage of graphene-based plasmonic couplers is their tunability. As mentioned before, the conductivity depends on various parameters such as electric bias, chemical doping, and magnetic bias. Therefore, tunable graphene-based couplers can be designed for various applications. Another interesting feature of graphene-couplers is their good confinement in the mid-infrared region, which can be utilized in designing compact nano-scale devices.

However, graphene couplers suffer from the large propagation loss, due to the lossy feature of the graphene layer. This disadvantage makes the implementation of couplers impractical. To overcome this, graphene technology can be integrated with other high-index contrast couplers such as the SOI platform. It must be emphasized that the fabrication of graphene-based structures has standard processes in commercial and many chemists work on the various methods of graphene fabrication.

Graphene couplers have many useful applications in mid-infrared and THz technologies. They can be utilized as an optical switch, plasmonic waveguide, polarization beam-splitter, plasmonic filter, circulator, magnetic probe, transparency windows, modulators, and unidirectional devices.

## 5. Conclusion

In this paper, a historical review of plasmonic graphene couplers was presented. Due to the importance of graphene conductivity in designing plasmonic couplers, the conductivity of graphene in presence of electrostatic or magnetostatic bias was considered at first. Then, various types of graphene-based couplers were introduced and their principal operations were studied more precisely. In all of these couplers except non-reciprocal couplers, the graphene was biased electrically while the anisotropic graphene had been utilized to propose non-reciprocal couplers. Graphene couplers exhibit some useful features such as tunability and good confinement in the mid-infrared region. They have many applications in the THz region such as plasmonic filter, the polarization beam splitter, plasmonic switch, etc.